# A High-Order Compensated Op-amp-less Bandgap Reference with 39 ppm/°C over -260~125 °C Temperature Range and -50 dB PSRR

Hechen Wang, *Student Member, IEEE*, Fa Foster Dai, *Fellow, IEEE*, and Michael Hamilton, *Member, IEEE*

*Abstract*—This brief presents a technique for compensating the temperature coefficient (TC) of a bandgap reference (BGR) using temperature characteristics of transistor's current gain *β*. As a comparison, three BGR circuits built with Si BJTs and SiGe HBTs are implemented to demonstrate the proposed TC curvature compensation technique. Measured average TC of the HBT proposed BGR is 23ppm/°C and 39 ppm/°C over the commercial (0~70°C) and space (-260~125°C) temperature ranges, respectively. With the proposed PSRR improvement technique, the BGR reaches PSRR of -50dB at 1MHz, and -38dB at 1GHz, respectively.

*Keywords*—temperature dependence of *β*, silicon, SiGe HBT, bandgap reference, curvature compensation, high order compensation.

## I. INTRODUCTION

Bandgap reference (BGR) circuits have been widely used in IC designs since they were introduced by Widlar [1]. Several BGRs using Widlar BGR concept or its variants have been reported in both bipolar and CMOS implementations.

Nowadays, BGRs are facing increasing challenges for ultra-wide temperature applications such as space applications and precise analog circuit designs such as high resolution data converters [2]-[4] and frequency synthesizers [5]. However, due to the curvature of the reference output voltage, the conventional Widlar BGR first-order compensated reference circuits exhibit limited temperature compensation, which is insufficient for wide temperature range applications. An efficient technique to compensate the curvature of BGR's *temperature coefficient* (TC, in unit of volts per degree K) using the current gain *β* of the BJTs [6] is adopted in our design, which significantly enlarges the range of compensation. Moreover, BGR circuits are normally affected by power supply variations. This work presents a simple, yet efficient, circuit that improves the performance of the BGR *power supply rejection ratio* (PSRR) from conventional BGR for more than 15 dB.

This brief investigates the temperature characteristics of *β* in both Si and SiGe processes and reports on the implementation of higher order curvature compensated BGRs utilizing its temperature characteristics. In Section II, the conventional Widlar BGR is introduced and its limitations are discussed. In Section III, *β* temperature dependences of Si and SiGe devices are analyzed. In Section IV, the proposed BGRs are presented. In Section V, measurement results are summarized.

## II. LIMITATION OF FIRST ORDER COMPENSATION

Figure 1 shows a basic BGR circuit. The first order temperature compensation is accomplished by the following equation:

$$V_{REF} = V_{BE} + \lambda V_T \quad (1)$$

where $V_{BE}$ is a negative TC parameter, $V_T$ is equal to $k_B T/q$, which is a positive TC parameter; and $\lambda$ is a scaling factor for optimizing the compensation. $k_B$ is Boltzmann factor and $q$ is electronic charge.

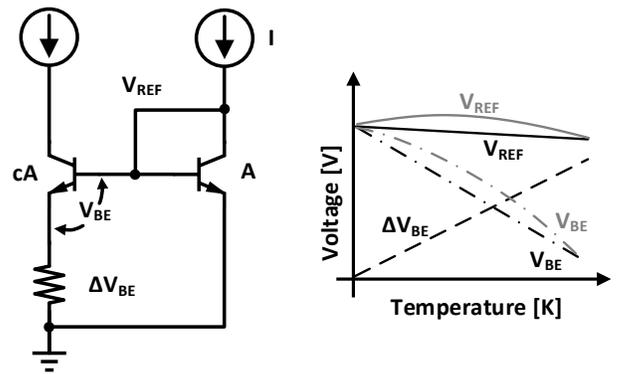

Fig. 1. A conventional BGR with the first order compensation.

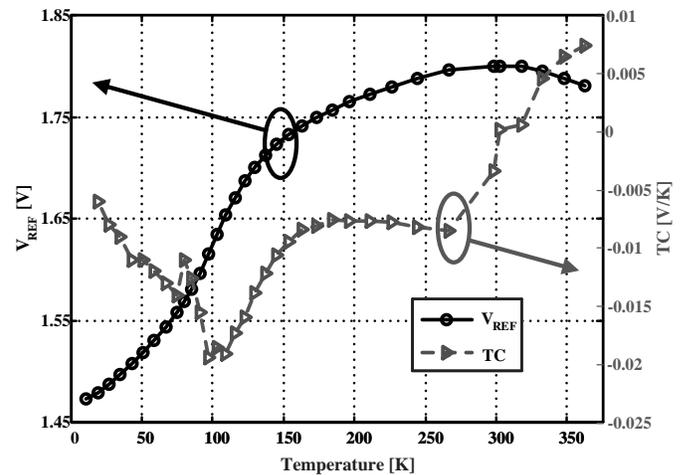

Fig. 2. Measured output and TC results of the first order BJT BGR.

The two transistor sizes in the circuit have a ratio of *c*, which leads to a slight difference $\Delta V_{BE}$. $\Delta V_{BE}$ is equal to $\ln(c) k_B T/q$. Then Eq. 1 can be rewritten as:

$$V_{REF} = V_{BE} + \ln(c)\frac{k_B T}{q}. \quad (2)$$

Assuming $V_{BE}$ is a linear term and by adjusting the ratio of *c* we can compensate the TC of the BGR independently such that its temperature compensated output $V_{REF}$ versus the temperature should be a horizontal straight line. However, the measurement results of a traditional first order BJT BGR shows that the slope of the BGR's output is nonzero versus temperature, as shown in Fig. 2.

The TC curve in Fig. 2 is the derivative of the BGR output $V_{REF}$. It shows that the slope of the output curve becomes zero only around 300 K. In a wide temperature range, the slope moves away from zero, namely the $V_{REF}$ is temperature independent only at 300 K, and the variation increases when the temperature offset increases. Obviously, parameter $V_{BE}$ contains higher order temperature dependent terms.

The TC of $V_{BE}$ has been studied extensively, such as in [7]. Considering the $V_{BE}$ temperature dependence, we can derive the $V_{REF}$, as:

$$V_{REF}(T) = V_{BE}(T) + \lambda V_T$$

$$V_{BE}(T) = V_G(T) + \left(\frac{T}{T_r}\right)[V_{BE}(T_r) - V_G(T_r)]$$

$$- (\mu(T)-1)\left(\frac{k_B T}{q}\right)\ln\left(\frac{T}{T_r}\right) + \lambda V_T \quad (3)$$

where $V_G$ is the bandgap voltage of silicon, $\mu$ is the charge carrier mobility, and $T_r$ is a reference temperature. It is evident from this expression that the $V_{BE}$ contains several temperature dependent parameters.

Mobility $\mu$ increases with increasing temperature in the low temperature regime because impurity scattering is reduced as the carriers gain more energy and have higher velocity at higher temperatures. However, if temperature further increases, $\mu$ will decrease due to lattice (phonon) scattering. A simulation result of mobility versus temperature for different doping levels is shown in Fig. 3. The results also show that the mobility $\mu$ and $d\mu/dT$ decreases with the doping level, which means heavily doped semiconductor devices can exhibit higher temperature stability.

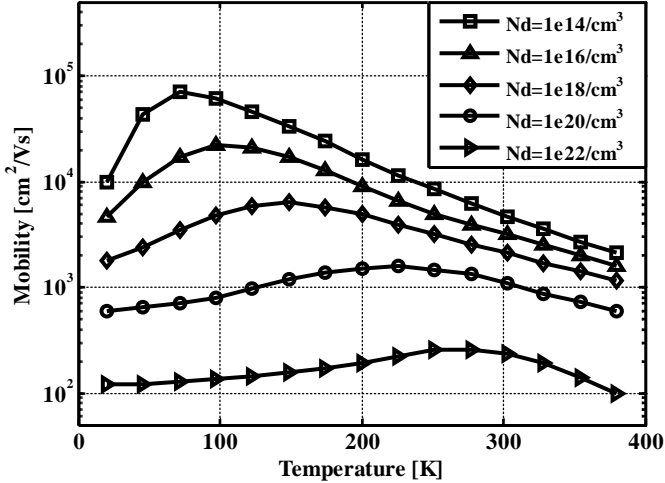

Fig. 3. Simulated mobility vs. temperature for different doping levels.

### III. TEMPERATURE DEPENDENCE OF BETA

The 1st order temperature compensation is widely adopted for commercial (0~70ºC) applications with low temperature fluctuations. However, in special fields, like military and space exploration, the temperature can vary between -200 to 125ºC, where high order compensation becomes indispensable.

There are several ways to cancel the higher order terms of $V_{BE}$. The approach adopted in this brief is to utilize the current gain $\beta$ of bipolar transistors. $\beta$ is a temperature dependent coefficient and is nonlinear, which can be used to compensate the higher order terms in $V_{BE}$.

Current gain $\beta$ can be expressed as follows, [8] [9]:

$$\beta = \frac{\alpha}{1-\alpha}, \quad \alpha = \frac{\text{sech}(W_B/L_{nb})}{1+(D_{pe}/D_{nb})(p_{ne}/n_{pb})(W_B/L_{pe})}. \quad (4)$$

$W_B$ is base width. $D_{pe}$ and $D_{pe}$ are diffusion constants of holes in the emitter and electrons in the base, namely:

$$D_{pe} = \mu_e \frac{k_B T}{q} \quad D_{nb} = \mu_b \frac{k_B T}{q}$$

$L_{pe}$ is hole path length in the emitter and $L_{nb}$ is electron path length in base. $p_{ne}/n_{pb}$ is minority carrier concentration ratio.

$$\frac{p_{ne}}{n_{pb}} = \frac{n_i^2/N_E}{n_i^2/N_B} = \frac{N_B}{N_E}\frac{N_{CE}N_{VE}\exp(-E_{gE}/k_B T)}{N_{CB}N_{VB}\exp(-E_{gB}/k_B T)}$$

$$= \frac{N_B}{N_E}\exp\left(\frac{E_{gB}-E_{gE}}{k_B T}\right)$$

where $N_{CE}$, $N_{VE}$ and $N_{CB}$, $N_{VB}$ are effective densities of states in conduction and valence bands of emitter and base semiconductors, respectively. $N_B$ and $N_E$ are base and emitter doping level. $n_i^2$ is proportional to $T^3 e^{-(E_g - \Delta E_g/k_B T)}$.

Substituting the above parameters into Eq. 4, we obtain the relationship between $\beta$ and $T$ as follows:

$$\beta \approx \frac{n_{pb}}{p_{ne}} = \frac{N_E}{N_B}\exp\left(\frac{E_{gE}-E_{gB}}{k_B T}\right)$$

Let $\beta_\infty = N_E/N_B$ and $\Delta E_g = E_{gE} - E_{gB}$, we get

$$\beta(T) = \beta_\infty \exp\left(\frac{\Delta E_g}{k_B T}\right) \quad (5)$$

The current gain $\beta$ is a temperature dependent parameter that follows an exponential rule, namely it contains higher order TC terms, according to Eq. 5. By taking a Taylor series expansion we find:

$$\beta(T) = c_0 - c_1(T-T_r) + c_2(T-T_r)^2 + \ldots + (-1)^n c_n (T-T_r)^n$$

The Taylor series coefficients $c_0$ $c_1$ … $c_n$ can be adjusted by changing the circuit's parameters and partially compensate the higher order terms in Eq. 2.

Furthermore, $\beta$ is also affected by $\Delta E_G$; that is, different types of material will lead to different temperature characteristics. This is the reason that Si based BJTs and SiGe based HBTs have opposite temperature characteristics.

To demonstrate BJT and HBT's $\beta$ temperature dependence characteristics, two types of BGRs, Si based BJT and SiGe based HBT, were built, simulated and tested on a 0.18μm SiGe BiCMOS process. The $\beta$ versus 1/temperature relationships of Si and SiGe transistors are shown in Fig. 4 and Fig. 5, respectively. The simulation is based on Cadence virtuoso tool and foundry provided data which only cover -25~125ºC temperature range. The simulation result is very close to calculation in its valid region. Measurement result shows a great similarity comparing with calculation as well.

The measurement results are very close to the simulations and calculations based on the theory. The polarity of BJT's and HBT's $\Delta E_G$ are different, according to Eq. 5. By tracing back from the measurement results in Fig. 4, the $\beta_\infty$ and $\Delta E_G$ of Si

BJT can be calculated, which are approximately 70 and -25m eV, respectively. On the contrary, SiGe HBT's $\beta_\infty$ and $\Delta E_G$ are obtained from Fig. 5, where we find $\beta_\infty$ is ~ 50, and $\Delta E_G$ is approximately 42m eV.

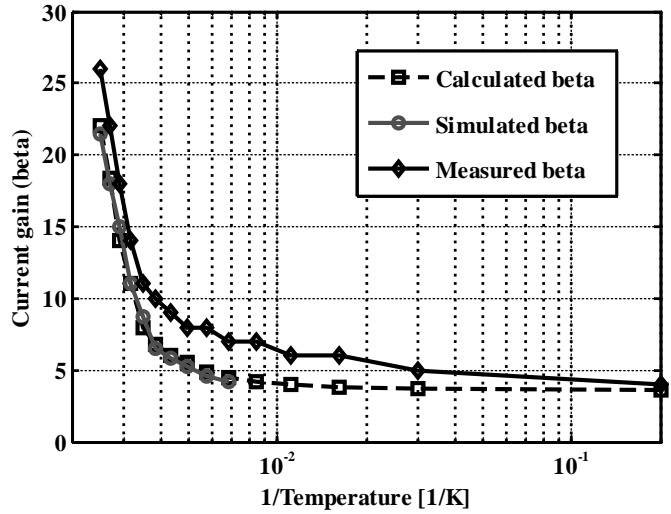

Fig. 4. Si BJT current gain $\beta$ versus 1/temperature.

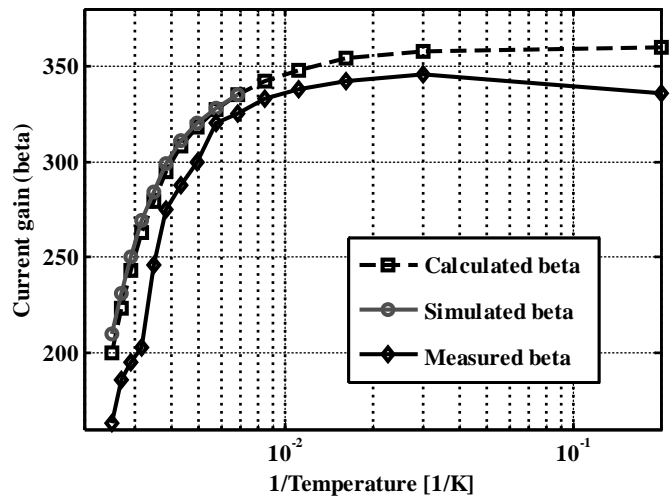

Fig. 5. SiGe HBT current gain $\beta$ versus 1/temperature.

The differences between HBT and BJT is that the HBT's base is built with heterojunction semiconductor material, which means the bandgap of the base is different from emitter and collector. HBT's $\Delta E_G$ is larger than BJT. Recall in equation (12), the larger $\Delta E_G$ is, the higher $\beta$ will be. It is for this reason that HBT has much higher current gain than Si BJT and has different temperature dependence characteristic, which means it is much easier to compensate the nonlinear terms in $V_{BE}$.

## IV. CURVATURE COMPENSATION

The crucial point of the high order temperature compensation is to utilize $\beta$'s TC to compensate the terms in $V_{BE}$. Fig. 6 illustrates a simple circuit that extracts the $\beta$ TC into the output voltage $V_{REF}$.

Two current sources $I_1$ and $I_2$ are proportional to absolute temperature (PTAT) sources, therefore the reference voltage $V_{REF}$ is given by

$$V_{REF}(T) \approx V_{BE}(T) + a_1 RT + \frac{a_2 RT}{\beta(T)}$$
$$= V_{BE}(T) + \lambda_1 T + \lambda_2 T \exp\left(\frac{\Delta E_G}{k_B T}\right) \quad (6)$$

In summary, the proposed curvature compensation circuit is given in Fig. 7. The ratio of two current sources $I_1$ and $I_2$ is related to the two parameters $\lambda_1$ and $\lambda_2$ in equation (13), which need to match the corresponding terms in $V_{BE}$.

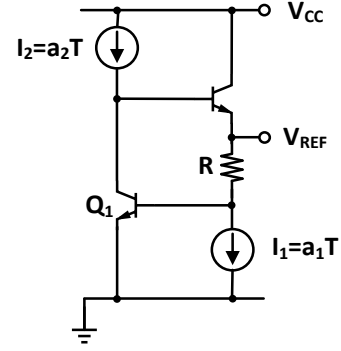

Fig. 6. Basic curvature compensation circuit.

Transistors $Q_1$ and $Q_2$ form the circuit illustrated in Fig. 6 and the two PTAT current sources are formed by two transistor pairs ($Q_3$, $Q_4$) and ($Q_5$, $Q_6$). Transistor $Q_7$ and resistor $R_5$ are solving for start-up issue and setting an initial current $I_{ini}$. The ratio of transistors $M_1$ and $M_2$ needs to be fine-tuned to adjust the $\beta$'s influence, which relates to inflection points and curvature of the output voltage.

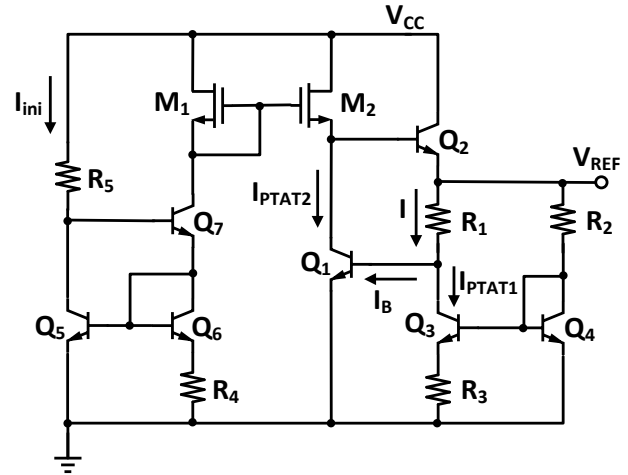

Fig. 7. Proposed curvature compensation circuit.

## V. PSRR IMPROVEMENT TECHNIQUE

The BGR circuit is powered by power regulators or directly tied to the power supply, which is sensitive to power supply variations. The output of the power supply can vary significantly and may couple unwanted signal or noise, both of which may affect the output of BGR.

Conventional BGR designs often use cascode transistors to separate the circuit from power supply, which provides some power supply rejection. However, cascode structures consume more headroom, which is not suitable for low power, low voltage applications. We propose a simple way to reduce the

PSRR without adding cascode transistors.

The proposed BGR circuit with PSRR improvement and curvature compensation techniques is shown in Fig. 8. The previous structure's initial current $I_{ini}$ is set by resistor $R_5$. $I_{ini}$ is equal to $\left[V_{CC}-\left(V_{BE,Q_7}+V_{BE,Q_5}\right)\right]/R_5$. Any disturbance on the power supply will directly affect the PTAT current source and cause BGR output fluctuation. By replacing $R_5$, in Fig. 7, with a power supply independent current source, one can greatly reduce the PSRR.

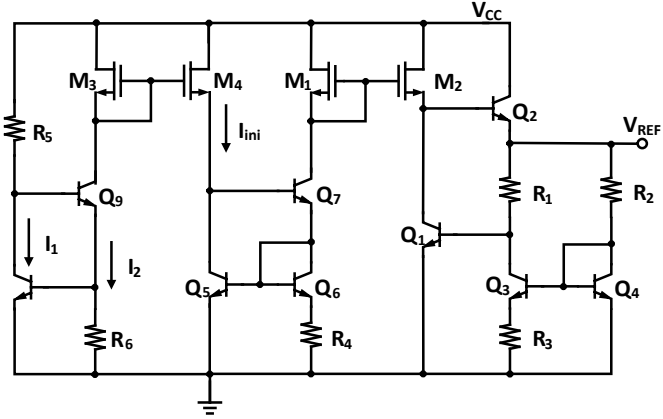

Fig. 8. Proposed BGR circuit with PSRR improvement technique.

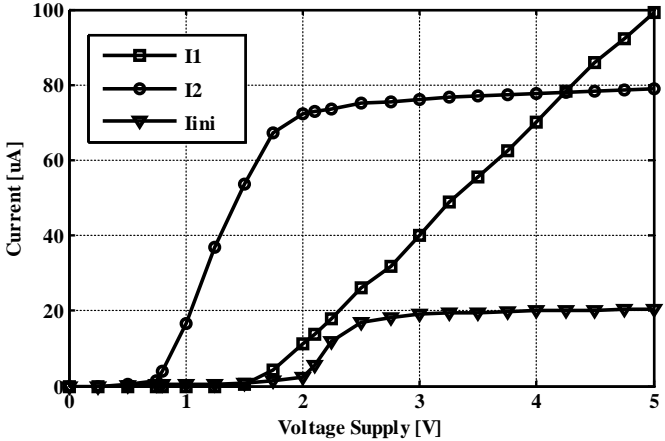

Fig. 9. BGR currents temperature dependence simulation results.

With the added circuit elements, the current $I_{ini}$ is equal to a ratio of $I_2$, which is generated by $V_{BE}$ of transistor $Q_8$ divided by $R_6$. Although $V_{BE}$ is a term related to $I_1$, the influence has been suppressed greatly. Furthermore, introducing a ratio between $M_3$ and $M_4$ can provide additional PSRR to the circuit. Fig. 9 shows the power supply dependence of currents $I_1$, $I_2$ and $I_{ini}$. From $I_1$ to $I_{ini}$, the power supply's influence has been reduced by 30 dB. The proposed technique provides a constant current to the BGR by using $I_C$-$V_{BE}$ relationship and the ratio of a current mirror. Theoretically, this structure can improve the PSRR by more than 30 dB without involving any op-amps.

## VI. MEASUREMENT RESULTS

Die photo of the proposed BGRs chip is shown in Fig. 12. Three different types of BGR (1st order compensation, 2nd order Si BJT and 2nd order SiGe HBT) were fabricated and tested. Each of them consumes 70x65 um$^2$ on-chip area and 94 uA average current.

The trace with circle marker in Fig. 10 illustrates measured 2nd order HBT BGR output voltage $V_{REF}$ versus temperature for the proposed circuit. It can be seen that it is not a parabolic curve anymore. The first order derivative TC crosses x axis twice. It has two zeros and one pole, which means this BGR's TC equals to zero at two points and significantly reduces the average TC.

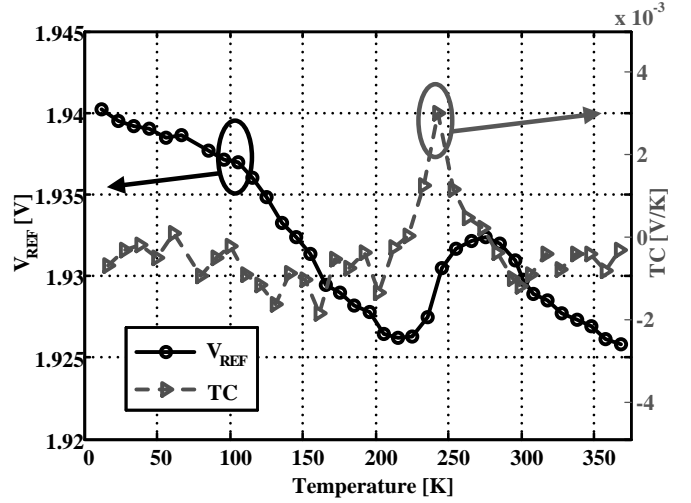

Fig. 10. Measured SiGe HBT based BGR output and TC curves.

The average TC of the SiGe HBT BGR is measured as 23 and 39 ppm/°C over the commercial (0 ~ 70 °C) and space (-260 ~ 125 °C) temperature ranges, respectively.

This work also adds a very simple circuit to 2nd order HBT BGR that improves the performance of PSRR by more than 15 dB compared with the other two BGRs. Since we did not use any op-amps in this circuit, the PSRR remains at very low level even dealing with very high frequency. 2nd order HBT BGR measurement results show that the power supply rejection ratios are -64 dB under 3.0 volt supply voltage at 1 MHz, and -42 dB at 1 GHz, respectively. Fig. 11 shows the PSRR versus frequency.

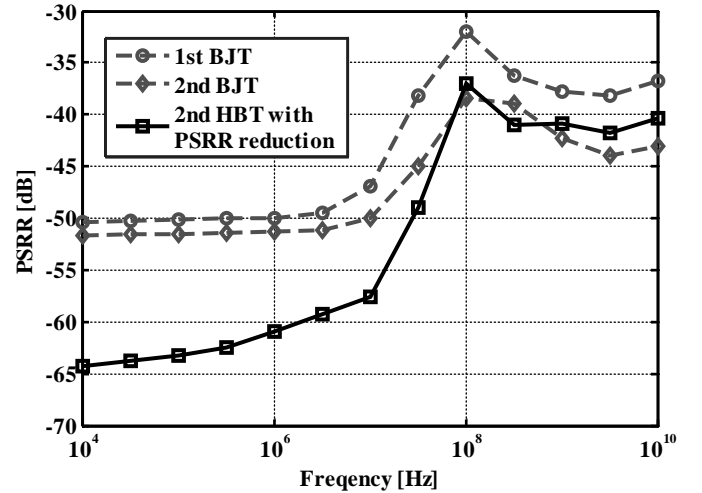

Fig. 11 Measured SiGe HBT based BGR output PSRR.

A Si BJT BGR version was built using CMOS transistors' parasitic bipolar transistors in BiCMOS technology. The compensation effect has also been demonstrated. The average

TC of the Si BJT BGR are measured as 63 and 133 ppm/ºC over the commercial and space temperature ranges, respectively. The PSRR are -64 dB under 3.0 volt supply voltage.

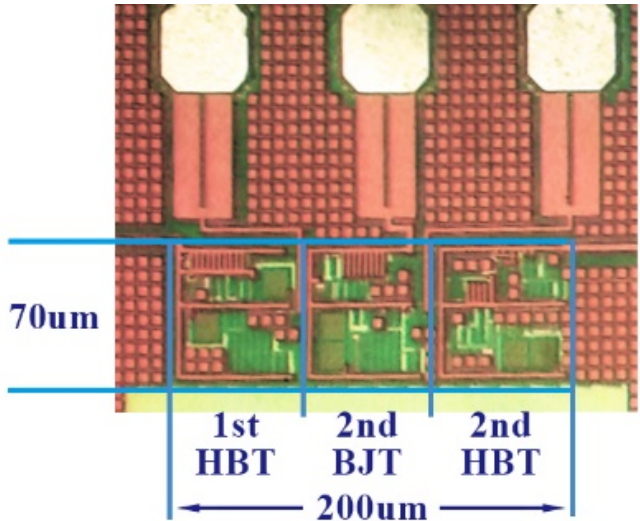

Fig. 12. Die photo of three different BGR circuits implemented for comparison.

TABLE I.  BGR PERFORMANCE SUMMARY*

| BGR Types | $V_{CC}$ [V] | TC [ppm/ºC] | PSRR [dB] | Power [mW] | Area [um²] |
|---|---|---|---|---|---|
| 1st BJT (CMOS) | 3.0 | 244 [1] | -50 [3] | 0.7 | 70x60 |
|  |  | 648 [2] | -38 [4] |  |  |
| 2nd BJT (CMOS) | 3.0 | 63 [1] | -52 [3] | 0.3 | 70x60 |
|  |  | 133 [2] | -43 [4] |  |  |
| 2nd HBT | 3.0 | 23 [1] | -64 [3] | 0.3 | 70x80 |
|  |  | 39 [2] | -42 [4] |  |  |

[1]. Over 0~70°C.  [2]. Over -260~125°C.  [3]. At 1MHz.  [4]. At 1GHz.
* All measurement results are obtained without trimming.

## VII. CONCLUSION

In this work, we analyzed the temperature dependence of bipolar transistor's $V_{BE}$, and discussed the deficiency of the 1st order compensated BGR due to the $V_{BE}$ temperature dependence. We then utilize the curvature cancellation technique by means of BJT or HBT's current gain $\beta$. The temperature dependence of $\beta$ was derived. Three BGRs are implemented using Si BJT and SiGe HBT in this work, and the measurement results demonstrate that even with different temperature dependence of the $\beta$, i.e., positive or negative, as long as it contains higher order terms, it can be used to compensate the 1st order BGR's TC, and thus improve the BGR's temperature compensation performance. Our HBT-based BGR shows better temperature independency than BJT BGR. A PSRR improvement circuit is developed for the 2nd order compensation BGR, which improves the PSRR performance by more than 30dB.

## ACKNOWLEDGMENT

We would like to thank Dr. Guofu Niu from Auburn University for many valuable discussions.

TABLE II.  BGRs PERFORMANCE COMPARISON

|  | TCASII 2010 [10] | TCASII 2010 [11] | TCASII 2015 [12] | This work |
|---|---|---|---|---|
| Temp. range (ºC) | 0~100 | -40~130 | -40~85 | **-260~125** |
| Supply voltage (V) | 3.0 | 3.6 | 1.4~3.6 | **2.2~5.0** |
| Output voltage (V) | 0.65 | 1.23 | 0.8 | **1.93** |
| TC (ppm/ºC) | 10.4 | 11.8 | 41.5 | **23/39*** |
| PSRR (dB) | -51 | -32 | / | **-64** |
| Power consumption (mW) | 0.14 | 0.6 | 0.06 | **0.3** |
| Area (mm²) | 0.011 | 0.1 | 0.009 | **0.006** |
| Process | 0.25 um CMOS | 0.5 um CMOS | 40 nm CMOS | **0.13 um SiGe** |